\documentclass[runningheads]{llncs}
\usepackage[T1]{fontenc}
\usepackage{graphicx}
\usepackage{multirow}
\usepackage{amsmath}
\usepackage{amsfonts}
\usepackage{xcolor}
\usepackage{booktabs}
\usepackage{etoolbox}
\usepackage{authblk}
\usepackage{fontawesome} 
\definecolor{mygreen}{RGB}{114, 240, 126}
\newcommand{\equalcontrib}{\textsuperscript{*}}
\newcommand{\Envelope}{\textsuperscript{\faEnvelopeO}}

\begin{document}

\title{Cardiac Copilot: Automatic Probe Guidance for Echocardiography with World Model}
\titlerunning{Automatic Probe Guidance for Echocardiography with World Model}

\author{
Haojun Jiang\inst{1,2}\equalcontrib \and
Zhenguo Sun\inst{2}\equalcontrib \and
Ning Jia\inst{2}\and
Meng Li\inst{2}\and
Yu Sun\inst{2}\and
Shaqi Luo\inst{2}\and
Shiji Song\inst{1}\and
Gao Huang\inst{1,2}\Envelope
}

\authorrunning{H. Jiang et al.}

\institute{Department of Automation, BNRist, Tsinghua University, Beijing, China\\
\and
Beijing Academy of Artificial Intelligence, Beijing, China\\
\email{jhj20@mails.tsinghua.edu.cn}, \email{hitsunzhenguo@gmail.com}, \email{gaohuang@tsinghua.edu.cn}
}

\maketitle
\renewcommand{\thefootnote}{}
\footnotetext[1]{\equalcontrib These authors contributed equally to this work. This work was done while Haojun Jiang was an intern at Beijing Academy of Artificial Intelligence.}
\footnotetext[2]{\Envelope Corresponding author.}

\begin{abstract}
Echocardiography is the only technique capable of real-time imaging of the heart and is vital for diagnosing the majority of cardiac diseases.
However, there is a severe shortage of experienced cardiac sonographers, due to the heart's complex structure and significant operational challenges.
To mitigate this situation, we present a Cardiac Copilot system capable of providing real-time probe movement guidance to assist less experienced sonographers in conducting freehand echocardiography.
This system can enable non-experts, especially in primary departments and medically underserved areas, to perform cardiac ultrasound examinations, potentially improving global healthcare delivery.
The core innovation lies in proposing a data-driven world model, named Cardiac Dreamer, for representing cardiac spatial structures.
This world model can provide structure features of any cardiac planes around the current probe position in the latent space, serving as an precise navigation map for autonomous plane localization.
We train our model with real-world ultrasound data and corresponding probe motion from 110 routine clinical scans with 151K sample pairs by three certified sonographers.
Evaluations on three standard planes with 37K sample pairs demonstrate that the world model can reduce navigation errors by up to 33\% and exhibit more stable performance.
\end{abstract}

\section{Introduction}
Cardiovascular diseases, as the primary cause of death and health risks in humans, affected 422.7 million people globally and led to 17.9 million deaths, accounting for 31\% of all global fatalities in 2015~\cite{roth2017global,song2020global}.
When screening for heart diseases, echocardiography is indispensable, as it is the only imaging technique that allows real-time observation of the heart's structure and function.
However, cardiac ultrasound is challenging to operate due to the following reasons: (1) the heart's complex structure; (2) limited observation windows due to rib obstruction; (3) the need for very precise adjustments; (4) particularly significant individual variations.
Consequently, echocardiography require highly skilled sonographers who possess extensive anatomical knowledge and rich operational experience.
According to the guideline of the American Society of Echocardiography~\cite{ehler2001guidelines,gardner1992guidelines}, an trainee with a medical background requires at least 12 months of clinical training before potentially being qualified for the position.
The long training period combined with the rising incidence of cardiovascular diseases has resulted in a significant shortage of skilled cardiac sonographers, especially pronounced in primary departments and medical underserved areas.

\begin{figure}[!tp]%
\centering
\includegraphics[width=\textwidth]{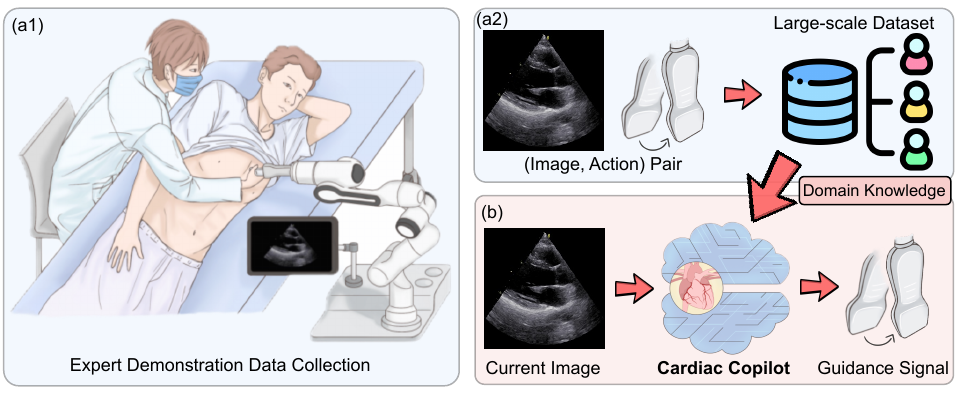}
\caption{
\textbf{Schematic diagram of cardiac copilot system.}
\textbf{(a1-2)} We collect a large-scale expert demonstration dataset with extensive ultrasound image and probe motion sample pairs using a robotic arm.
The demonstration data contains critical medical knowledge required for performing echocardiography.
\textbf{(b)} We encapsulate experts’ knowledge within deep neural networks to facilitate automatic probe guidance.
}
\label{fig1}
\end{figure}

With technological advancements~\cite{ha2018world,he2016deep,hussein2017imitation,jiang2022pseudo,jiang2022cross,yang2021condensenet}, AI-driven systems show immense potential to significantly improve the efficiency of medical ultrasound processes.
For example, Droste~\textit{et al.}~\cite{droste2020automatic} proposed an US-GuideNet to provide only rotation movement guidance signal for acquiring standard planes in routine freehand obstetric ultrasound scanning.
However, in routine ultrasound scanning, the first step often involves moving close to the target plane through translation.
Thus, providing a six-dimension (6D) guidance signal is essential, which includes signals for both translation and rotation.
Narang \textit{et al.}~\cite{narang2021utility} proposed an AI-driven commercial solution to guide nurses to acquire echocardiogram by suggesting approximate adjustment directions in one degree of freedom.
They further conducted clinical trials and found that it could assist nurses in locating cardiac target planes that meet diagnostic requirements.
Nonetheless, the specifics of the methodological implementation remain undisclosed, impeding progress in cardiac ultrasound domain.
Later, Shida \textit{et al.}~\cite{shida2023automated,shida2023diagnostic} investigated the potential of using manually predefined search strategies to acquire the Parasternal Long-Axis (PLAX) plane.
The authors conducted a small-scale experiment on five healthy subjects, with the upper limit of the success rate being only 63.3\%. 
This indicates that the rule-based guidance strategy lacks generalizability and practicality.
Despite on-going eﬀorts, there is still no publicly available research on cardiac probe navigation, which hampers further advancement in the field.

In this paper, we present an AI-driven cardiac probe guidance system, named Cardiac Copilot, capable of providing real-time probe guidance signals for locating target standard planes in routine freehand echocardiography.
As illustrated in Fig.~\ref{fig1}, the fundamental logic involves an artificial intelligence network that, based on the current ultrasound image, outputs a six-dimension action command to guide the operator towards the target standard plane.
In clinical practice, highly skilled sonographers can locate the target section with fewer explorations, as they can imagine the potential image that might result from a specific adjustment based on the current image, thereby achieving accurate adjustment.
Inspired by the above domain expertise, we propose a data-driven world model named Cardiac Dreamer to represent the spatial structure of the heart.
Specifically, it can ``imagine'' the features of adjacent continuous planes in the latent space based on the current ultrasound image and relative positional relationships.
Upon integrating Cardiac Dreamer into the system, it effectively gains a precise map in latent space, significantly enhancing its navigational proficiency.
We collected expert operation data on acquiring standard planes from 125 routine clinical scans with 188K sample pairs by three certified sonographer, to drive the model's learning of generalizable guidance capabilities.
The data were collected by manually operating a probe attached to the end of a robotic arm, with precise probe 6D pose information obtained through the arm's internal sensors.
Assessments conducted on three standard planes using 37K pairs of samples reveal that the world model can decrease navigation errors by up to 33\% and demonstrate enhanced stability in performance.

\section{Method}
In this section, we provide details regarding the methodological implementation about automatic probe guidance for freehand echocardiography.
We first introduce the target-oriented guidance framework which enables the base capabilities of locating standard planes in Sec \ref{sec:vanilla_guidance}.
Moreover, in Sec \ref{sec:cardiac_dreamer}, we present the Cardiac Dreamer, which foresees the state that would result from the executing predicted action.
By collaborating with Cardiac Dreamer, the base model is able to exhibit improved guidance performance.

\subsection{Target-oriented Guidance Framework}
\label{sec:vanilla_guidance}
In clinical practice, the ultimate goal of cardiac ultrasound is to locate each standard two-dimensional plane and observe the heart's structure and function on these standard planes.
Consequently, we aim to train a guidance network whose goal is to directly predict the relative positional relationship between the current probe location and the target plane, without the need to consider the intermediate path, thereby making it target-oriented.
Achieving this functionality is challenging as it requires the network to understand cardiac anatomy, interpret ultrasound images, and have spatial imagination skills to map 2D ultrasound images onto 3D cardiac structures.
In fact, during routine examinations, sonographers' decisions on each probe adjustment is the result of comprehensively applying these knowledge and skills.
Thus, to acquire guidance capabilities, we adopt the data-driven imitation learning strategy~\cite{duan2017one,ho2016generative,hussein2017imitation} to learn extensive domain knowledge from the large-scale expert demonstration dataset.

\begin{figure}[!tp]%
\centering
\includegraphics[width=\textwidth]{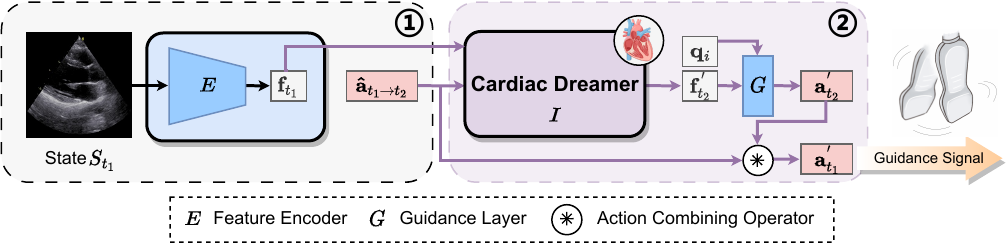}
\caption{
\textbf{Diagram of target-oriented guidance framework.}
\textit{Left} is the policy network which provides the basic guidance signal for locating the target plane.
\textit{Right} is the Cardiac Dreamer which foresees states reached by executing actions output by the policy network and refines the actions based on these states.
The details of the action combining operator is shown in Eq.~\ref{eq:aciton_combine}.
}
\label{fig2}
\end{figure}

Given an input ultrasound image at time $t$, representing the current state $\mathbf{S}_t \in \mathbb{R}^{H \times W}$, the operator needs a 6D guidance signal $\mathbf{a}_t \in \mathbb{R}^{6}$ to locate the target plane.
This six-dimensional vector $\mathbf{a}_t$ comprises the first three dimensions representing translational movements in the x, y, and z directions, while the last three dimensions indicate rotational movements in roll, pitch, and yaw.
Let $\mathcal{D} = \{d_1, d_2, \cdots, d_n\}$ be all the expert demonstration data collected from the routine examinations.
Specifically, $d_i = \{(\mathbf{S}_t, \mathbf{p}_t) | \mathbf{S}_t \in \mathbb{R}^{H \times W}, \mathbf{p}_t \in \mathbb{R}^{6}, t \in \mathcal{T}\}$ is one demonstration of standard plane acquisition, where $\mathbf{p}_t$ is 6D probe's pose information and corresponding ultrasound image $\mathbf{S}_t$ at time $t$.
When \( t = T \), \( \mathbf{S}_T \) is the standard section and \( p_T \) is the probe's target pose. 
By comparing other poses \( p_t \) in the demonstration sequence with \( p_T \), we obtain the $\mathcal{D}^{'}$ with each sequence $d_{i}^{'}$ as $ \{(\mathbf{S}_t, \mathbf{\hat{a}}_t) | t \in \mathcal{T} \setminus {T} \}$ with groundtruth supervision $\mathbf{\hat{a}}_t$.
Consequently, we train a policy network $\pi_{\theta}: \mathbf{S}_t \mapsto \mathbf{a}_t$ with parameters $\theta$.
As shown in Fig.~\ref{fig2}, this network consists of a feature encoder $E(\cdot)$ and a guidance layer $G(\cdot)$:
\begin{align}
    \mathbf{a}_t &= G(E(\mathbf{S_t}), \mathbf{q}_i),
\end{align}
where $\mathbf{q}_i, i\in[0, Q-1]$ is a learnable query feature, indicating one of the target planes.
Finally, the policy network $\pi_{\theta}$ is optimized with the following loss:
\begin{align}
    \mathcal{L}_{total}&= \frac{1}{|\mathcal{D}^{'}|} \sum_{d_i^{'} \in \mathcal{D}^{'}} \frac{1}{|d_i^{'}|} \sum_{(\mathbf{S_t}, \mathbf{\hat{a}}_t) \in d_i^{'}} \mathcal{L}_{\mathrm{SmoothL1}}(\mathbf{a_t}, \mathbf{\hat{a}_t}),
\end{align}
where $|\mathcal{D}|$ and $|D_i|$ are the number of elements inside each set.

\subsection{Foresee States by Cardiac Dreamer}
\label{sec:cardiac_dreamer}
This task can be likened to an autonomous navigation mission with the probe near the chest area, where it is usually necessary to construct an precise dense map to understand the relative positional relationships between any points.
However, the aforementioned network only utilizes sparse relative positional information between arbitrary planes and the target plane.
Upon closer examination of the demonstration data, we find a wealth of unused relative positional information between non-target planes.
Assuming there are $M$ intermediate planes in a demonstration sequence, the unused structural information accounts for $(M-1)/M$ of the total data.
Therefore, based on the extensive positional relationships, we can construct a world model~\cite{ha2018world,hafner2019dream,hafner2019learning} for heart, named Cardiac Dreamer, representing dense cardiac structures in latent space.

Given the processed demonstration dataset $D_{i}^{'}$, we create the permutation dataset $D^{''}$ with $d_{i}^{''} = \{(\mathbf{S}_{t_1}, \mathbf{S}_{t_2}, \mathbf{\hat{a}}_{t_1 \to t_2}, \mathbf{\hat{a}}_{t_1}, \mathbf{\hat{a}}_{t_2}) | t_1 \neq t_2 \}$, where $\mathbf{\hat{a}}_{t_1 \to t_2}$ is relative positional between $S_{t_1}$ and $S_{t_2}$, and $\mathbf{\hat{a}}_{t_1}$, $\mathbf{\hat{a}}_{t_2}$ is relative positional between $S_{t_1}$, $S_{t_2}$ and the target plane.
Then, $\mathbf{S}_{t_1}$ is processed by the feature encoder $E(\cdot)$ inside the policy network to yield $\mathbf{f}_{t_1}$:
\begin{align}
    \mathbf{f}_{t_1} = E(\mathbf{S}_{t_1}).
\end{align}
Cardiac Dreamer $I(\cdot)$ receives $\mathbf{f}_{t_1}$ and $\mathbf{\hat{a}}_{t_1 \to t_2}$ as input and outputs the feature $\mathbf{f}_{t_2}^{'}$ of the corresponding plane:
\begin{align}
    \mathbf{f}_{t_2}^{'} = I(\mathbf{f}_{t_1}, \mathbf{\hat{a}}_{t_1 \to t_2}).
\end{align}
We aim for Cardiac Dreamer to learn the spatial relationships between various planes, thereby gaining a comprehensive understanding of cardiac structures.
Furthermore, the generated $\mathbf{f}_{t_2}^{'}$ is used to predict the guidance signal $\mathbf{a}_{t_2}^{'}$ for locating the target plane:
\begin{align}
    \mathbf{a}_{t_2}^{'} = G(\mathbf{f}_{t_2}^{'}, \mathbf{q}_i).
\end{align}
Then, the predicted guidance signal $\mathbf{a}_{t_2}^{'}$ is combined with $\mathbf{\hat{a}}_{t_1 \to t_2}$ to get $\mathbf{a}_{t_1}^{'}$:
\begin{align}\label{eq:aciton_combine}
    \mathbf{a}_{t_1}^{'} = U^{-1}(U(\mathbf{\hat{a}}_{t_1 \to t_2}) U(\mathbf{a}_{t_2}^{'})),
\end{align}
where \(U(\cdot)\) is the function that converts the vector-form action \(\mathbf{a}\) into the standard homogeneous transformation matrix \(U(\mathbf{a}) \in \mathbb{R}^{4 \times 4}\). 
Thus, the total loss for joint optimization of the policy network with a Cardiac Dreamer is:
\begin{align}
    \mathcal{L}_{total} = \mathcal{L}_{\mathrm{SmoothL1}}(\mathbf{a}_{t_1}^{'}, \mathbf{\hat{a}}_{t_1}) +  \mathcal{L}_{\mathrm{SmoothL1}}(\mathbf{a}_{t_2}^{'}, \mathbf{\hat{a}}_{t_2}).
\end{align}

In experiments, we observe that Cardiac Dreamer can improve the guidance performance and provide more stable guidance signal.

\section{Experiments}

\subsection{Setup}
In the experimental setup, three standard echocardiographic planes, \textit{i.e.}, the Parasternal Long-Axis (PLAX), the Parasternal Short-Axis Aortic Valve (PSAX-AV), and the Parasternal Short-Axis Mitral Valve (PSAX-MV), were chosen for their clinical importance, offering critical insights into cardiac anatomy and functionality.
The ultrasound image data was acquired from health adult male volunteers using a Franka Panda robot arm with an ultrasound probe (M5S probe, GE Vivid E7 machine) manipulated by professional echocardiographers, meanwhile the pose information of the probe was also recorded synchronously by utilizing Robot Operation System (ROS).
Overall, approximately 188K cardiac ultrasound images paired with probe motion were collected from 125 routine clinical scans, forming a diverse dataset for training the proposed Cardiac Copilot models.
The whole data collecting process was approved and supervised by The Tsinghua University Science and Technology Ethics Committe.
Data processing involved aligning relative pose coordinates of each image with target planes which were identified by professionals, ensuring a unified coordinate system for each plane. 
The dataset was split into 110 scans (about 151K images) for training and 15 scans (about 37K images) for testing.
Note that the test set included individuals not seen during training to evaluate the algorithm's generalization.

\subsection{Implementation Details}
As shown in Fig.~\ref{fig2}, the model architecture includes a ResNet-34 for feature extraction, a guidance layer with three fully connected layers (the output dimensions are 2048, 512, and 6), and a Cardiac Dreamer network based on the original transformer architecture proposed in \cite{attention}. Additionally, one-hot encoding for plane selection is introduced, allowing the model to adapt relative pose predictions based on the chosen target plane. 
Model training, conducted with Distributed Data Parallel (DDP) training on four NVIDIA RTX 3090 GPUs, spanned 300 epochs. 
The training process employed the CosineAnnealingLR scheduler with 0.0001 learning rate, AdamW optimizer, and Automatic Mixed Precision (AMP).
For inference, we tested on an Nvidia RTX 3090, and the inference time for a single image is 59.4ms which meets practical requirements.

\subsection{Evaluation Metrics} 
In order to assess the accuracy and robustness of our proposed Cardiac Dreamer compared to the policy network which comprises merely feature encoder and guidance layer as baseline, we designed the evaluation metrics deliberately.
The primary metric for evaluation is the Mean Absolute Error (MAE) between the predicted and ground truth probe poses. This quantitative measure assesses the accuracy of our model in imitating the spatial movement of the echocardiography probe. Simultaneously, the stability of the model was evaluated by measuring the standard deviation of the absolute error to the target plane for all poses from the evaluation scans.

\begin{table}[!t]\small
\caption{
\textbf{Comparisons between the cardiac dreamer and the baseline.} 
We report MAE results representing probe guidance errors across three standard planes.}
\label{tab:mae}
\begin{center}
\resizebox{1\columnwidth}{!}{%
\begin{tabular}{p{1.7cm} p{2.7cm} p{2.2cm} p{2.2cm} p{2.2cm} p{2.2cm} p{2.2cm} p{2.4cm}}
\toprule
\multirow{2}{*}{\textbf{Plane}} & \multirow{2}{*}{\textbf{Model}} & \multicolumn{3}{c}{\textbf{Translation (mm)}} & \multicolumn{3}{c}{\textbf{Rotation (degree)}}\\
\cline{3-8}
 & & \textbf{x} & \textbf{y} & \textbf{z} & \textbf{rx} & \textbf{ry} & \textbf{rz} \\
\midrule
\multirow{2}{*}{PLAX} & Baseline & 7.85 & 5.97 & 3.64 & 6.77 & 4.92 & 10.46 \\
\cline{2-8}
 & Cardiac Dreamer & \textbf{6.10} \textbf{\textcolor{mygreen}{(-22.3\%)}}  & \textbf{4.85} \textbf{\textcolor{mygreen}{(-18.8\%)}} & \textbf{2.64} \textbf{\textcolor{mygreen}{(-27.5\%)}} & \textbf{5.76} \textbf{\textcolor{mygreen}{(-14.9\%)}} & \textbf{4.46} \textbf{\textcolor{mygreen}{(-9.3\%)}} & \textbf{9.57} \textbf{\textcolor{mygreen}{(-8.5\%)}} \\
\midrule
\multirow{2}{*}{PSAX-AV} & Baseline & 7.70 & 6.01 & 3.30 & 5.17 & 7.00 & 13.26 \\
\cline{2-8}
 & Cardiac Dreamer & \textbf{7.48} \textbf{\textcolor{mygreen}{(-2.9\%)}} & \textbf{4.83} \textbf{\textcolor{mygreen}{(-19.6\%)}} & \textbf{2.67} \textbf{\textcolor{mygreen}{(-19.1\%)}} & \textbf{4.20} \textbf{\textcolor{mygreen}{(-18.8\%)}} & \textbf{5.65} \textbf{\textcolor{mygreen}{(-19.3\%)}} & \textbf{9.77} \textbf{\textcolor{mygreen}{(-26.3\%)}} \\
\midrule
\multirow{2}{*}{PSAX-MV} & Baseline & 10.11 & 8.21 & 4.45 & 5.50 & 8.57 & 13.20\\
\cline{2-8}
 & Cardiac Dreamer & \textbf{8.53} \textbf{\textcolor{mygreen}{(-15.6\%)}} & \textbf{6.93} \textbf{\textcolor{mygreen}{(-15.6\%)}} & \textbf{3.62} \textbf{\textcolor{mygreen}{(-18.7\%)}} & \textbf{4.57} \textbf{\textcolor{mygreen}{(-16.9\%)}} & \textbf{5.73} \textbf{\textcolor{mygreen}{(-33.1\%)}} & \textbf{10.21} \textbf{\textcolor{mygreen}{(-22.7\%)}} \\
\bottomrule
\end{tabular}
}
\end{center}
\end{table}

\begin{figure}[!t]%
\centering
\includegraphics[width=\textwidth]{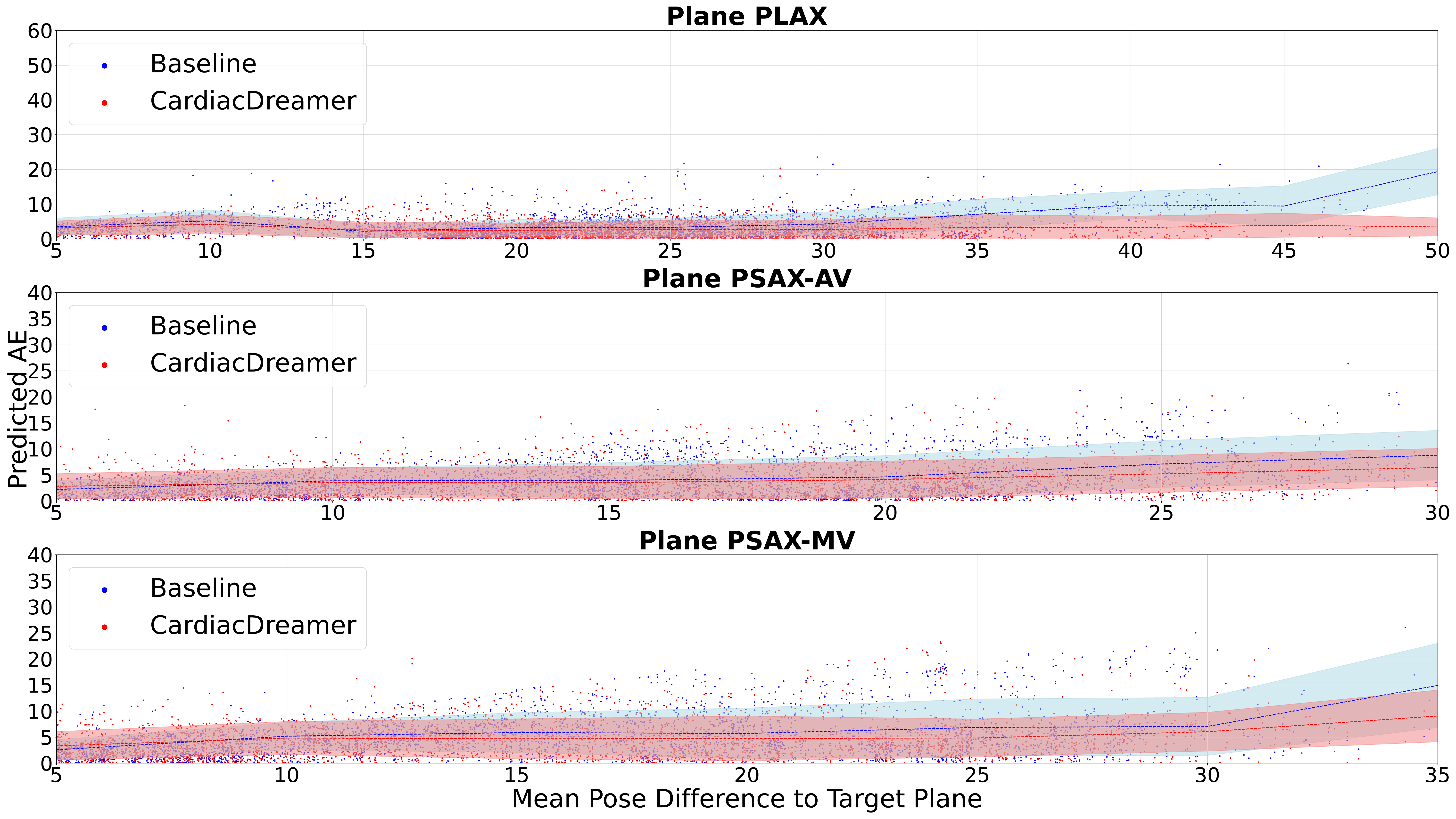}
\caption{
\textbf{Detailed analysis.}
For optimal viewing, please consider enlarging and viewing in color.
The change of predicted absolute error of different samples with respect to the mean pose difference to target plane on three standard planes.
The 'mean pose difference' refers to calculating the mean absolute error between the pose of the current plane and that of the target plane.
The dashed line represents the mean value, and the shading represents the standard deviation.
}
\label{fig:result}
\end{figure}

\subsection{Results and Analysis}
\paragraph{Accuracy.} The Table.~\ref{tab:mae} presents the MAE results for probe guidance prediction across three standard echocardiographic planes: PLAX, PSAX-AV, and PSAX-MV. 
The results indicate that the Cardiac Dreamer model consistently outperforms the baseline model across all translation and rotation components. 
The improvements are most pronounced in the z translation component for PLAX, rz rotation component for PSAX-AV, and ry rotation component for PSAX-MV, with 27.5\%, 26.3\%, and 33.1\% reduction in MAE respectively. 
These findings suggest that Cardiac Dreamer's ability to understand and model the spatial relationships and anatomical structures involved in echocardiographic imaging contributes to its enhanced guidance accuracy.

\paragraph{Stability.} The relationship between all probe poses and their Absolute Error (AE) to target planes is shown in Fig.~\ref{fig:result}. 
The figure shows, the Cardiac Dreamer model achieves significantly lower AE compared to the baseline model across different mean pose differences to the target plane. 
Moreover, the shaded regions for Cardiac Dreamer are narrower, indicating a smaller standard deviation and more consistent performance with different poses. 
The inclusion of standard deviation in the plots provides valuable insights into the reliability and robustness of the models' predictions. 
A smaller standard deviation, as observed for Cardiac Dreamer, suggests that the model's predictions are more tightly clustered around the mean, indicating greater precision and consistency across different scans. 
Conversely, a larger standard deviation, as seen for the baseline model, implies greater variability in the predictions, potentially indicating sensitivity to specific probe poses or scans. This variability could lead to less reliable performance in practical applications.

\paragraph{Analysis.} Overall, the results demonstrate that the proposed Cardiac Dreamer model not only achieves superior general performance in terms of navigation guidance accuracy but also exhibits greater consistency and robustness, as evidenced by the lower mean absolute error and the smaller standard deviations across all three echocardiographic planes evaluated. This combination of improved accuracy and reliability makes Cardiac Dreamer a more suitable choice for real-world scenarios in intelligent probe guidance scanning.

\section{Conclusion}
In this work, we present a Cardiac Copilot system designed to offer real-time guidance on probe movements, aiding novice sonographers in performing freehand echocardiography.
We innovatively introduce a data-driven world model named Cardiac Dreamer, which represents cardiac structures in feature space and serves as a latent map for automatic probe navigation.
Evaluations on the large-scale dataset show that Cardiac Dreamer consistently reduces the guidance errors across six-dimensions and exhibits more stable performance.
AI-driven probe guidance system shows the potential to reduce sonographer skill requirements, thereby increasing the capacity for cardiac ultrasound examinations in primary hospitals and areas with scarce medical resources.
In the future, the probe guidance system can also be used to direct a robotic arm for fully autonomous cardiac ultrasound scanning, serving as a core module in autonomous robots. 
This is a promising future direction that can significantly reduce repetitive manual operations, greatly enhance the supply capability of cardiac ultrasound, and improve the standardization of the examination process.

\subsubsection{Acknowledgement.}
This work was supported in part by the National Key R\&D Program of China (2021ZD0140407), the NSFC (62321005) and the Deng Feng Fund.

\subsubsection{Disclosure of Interests.}
The authors have no competing interests to declare that are relevant to the content of this article.

\bibliographystyle{splncs04}
\bibliography{Paper}

\begin{thebibliography}{10}
\providecommand{\url}[1]{\texttt{#1}}
\providecommand{\urlprefix}{URL }
\providecommand{\doi}[1]{https://doi.org/#1}

\bibitem{droste2020automatic}
Droste, R., Drukker, L., Papageorghiou, A.T., Noble, J.A.: Automatic probe movement guidance for freehand obstetric ultrasound. In: Medical Image Computing and Computer Assisted Intervention--MICCAI 2020: 23rd International Conference, Lima, Peru, October 4--8, 2020, Proceedings, Part III 23. pp. 583--592. Springer (2020)

\bibitem{duan2017one}
Duan, Y., Andrychowicz, M., Stadie, B., Jonathan~Ho, O., Schneider, J., Sutskever, I., Abbeel, P., Zaremba, W.: One-shot imitation learning. Advances in neural information processing systems  \textbf{30} (2017)

\bibitem{ehler2001guidelines}
Ehler, D., Carney, D.K., Dempsey, A.L., Rigling, R., Kraft, C., Witt, S.A., Kimball, T.R., Sisk, E.J., Geiser, E.A., Gresser, C.D., et~al.: Guidelines for cardiac sonographer education: recommendations of the american society of echocardiography sonographer training and education committee. Journal of the American society of echocardiography  \textbf{14}(1),  77--84 (2001)

\bibitem{gardner1992guidelines}
Gardner, C.J., Brown, S., Hagen-Ansert, S., Harrigan, P., Kisslo, J., Kisslo, K., Kwan, O.L., Menapace, F., Otto, C., Pandian, N., et~al.: Guidelines for cardiac sonographer education: report of the american society of echocardiography sonographer education and training committee. Journal of the American Society of Echocardiography  \textbf{5}(6),  635--639 (1992)

\bibitem{ha2018world}
Ha, D., Schmidhuber, J.: World models. arXiv preprint arXiv:1803.10122  (2018)

\bibitem{hafner2019dream}
Hafner, D., Lillicrap, T., Ba, J., Norouzi, M.: Dream to control: Learning behaviors by latent imagination. arXiv preprint arXiv:1912.01603  (2019)

\bibitem{hafner2019learning}
Hafner, D., Lillicrap, T., Fischer, I., Villegas, R., Ha, D., Lee, H., Davidson, J.: Learning latent dynamics for planning from pixels. In: International conference on machine learning. pp. 2555--2565. PMLR (2019)

\bibitem{he2016deep}
He, K., Zhang, X., Ren, S., Sun, J.: Deep residual learning for image recognition. In: Proceedings of the IEEE conference on computer vision and pattern recognition. pp. 770--778 (2016)

\bibitem{ho2016generative}
Ho, J., Ermon, S.: Generative adversarial imitation learning. Advances in neural information processing systems  \textbf{29} (2016)

\bibitem{hussein2017imitation}
Hussein, A., Gaber, M.M., Elyan, E., Jayne, C.: Imitation learning: A survey of learning methods. ACM Computing Surveys (CSUR)  \textbf{50}(2),  1--35 (2017)

\bibitem{jiang2022pseudo}
Jiang, H., Lin, Y., Han, D., Song, S., Huang, G.: Pseudo-q: Generating pseudo language queries for visual grounding. In: Proceedings of the IEEE/CVF Conference on Computer Vision and Pattern Recognition. pp. 15513--15523 (2022)

\bibitem{jiang2022cross}
Jiang, H., Zhang, J., Huang, R., Ge, C., Ni, Z., Lu, J., Zhou, J., Song, S., Huang, G.: Cross-modal adapter for text-video retrieval. arXiv preprint arXiv:2211.09623  (2022)

\bibitem{narang2021utility}
Narang, A., Bae, R., Hong, H., Thomas, Y., Surette, S., Cadieu, C., Chaudhry, A., Martin, R.P., McCarthy, P.M., Rubenson, D.S., et~al.: Utility of a deep-learning algorithm to guide novices to acquire echocardiograms for limited diagnostic use. JAMA cardiology  \textbf{6}(6),  624--632 (2021)

\bibitem{roth2017global}
Roth, G.A., Johnson, C., Abajobir, A., Abd-Allah, F., Abera, S.F., Abyu, G., Ahmed, M., Aksut, B., Alam, T., Alam, K., et~al.: Global, regional, and national burden of cardiovascular diseases for 10 causes, 1990 to 2015. Journal of the American college of cardiology  \textbf{70}(1),  1--25 (2017)

\bibitem{shida2023automated}
Shida, Y., Kumagai, S., Tsumura, R., Iwata, H.: Automated image acquisition of parasternal long-axis view with robotic echocardiography. IEEE Robotics and Automation Letters  (2023)

\bibitem{shida2023diagnostic}
Shida, Y., Sugawara, M., Tsumura, R., Chiba, H., Uejima, T., Iwata, H.: Diagnostic posture control system for seated-style echocardiography robot. International Journal of Computer Assisted Radiology and Surgery  \textbf{18}(5),  887--897 (2023)

\bibitem{song2020global}
Song, P., Fang, Z., Wang, H., Cai, Y., Rahimi, K., Zhu, Y., Fowkes, F.G.R., Fowkes, F.J., Rudan, I.: Global and regional prevalence, burden, and risk factors for carotid atherosclerosis: a systematic review, meta-analysis, and modelling study. The Lancet Global Health  \textbf{8}(5),  e721--e729 (2020)

\bibitem{attention}
Vaswani, A., Shazeer, N., Parmar, N., Uszkoreit, J., Jones, L., Gomez, A.N., Kaiser, L., Polosukhin, I.: Attention is all you need. In: Proceedings of the 31st International Conference on Neural Information Processing Systems. p. 6000–6010. Curran Associates Inc. (2017)

\bibitem{yang2021condensenet}
Yang, L., Jiang, H., Cai, R., Wang, Y., Song, S., Huang, G., Tian, Q.: Condensenet v2: Sparse feature reactivation for deep networks. In: Proceedings of the IEEE/CVF Conference on Computer Vision and Pattern Recognition. pp. 3569--3578 (2021)

\end{thebibliography}

\end{document}